\documentclass{article}       

\begin{document}

\title{Analysis of the particle transfer between two systems under unification}
\author{I.A. Molotkov$^1$,  A.I. Osin$^1$ \\
\\
$^1$ Institute of Terrestrial Magnetism, the Ionosphere\\
and Radio Wave Propagation (IZMIRAN), Moscow, Troitsk, Russia \\
e-mail: iamolotkov@yandex.ru, osin@izmiran.ru
}
\date{\today}
\maketitle

\begin{abstract}
We investigate unification of two systems of identical elements having different dimensions which may be of interest for both physics and economics.
Characteristic parameters as well as explicit formulae for the temperature (in economics - capital turnover) and dimension of the united system are obtained as functions of parameters of the initial systems.  Expressions also found for the entropies of initial and united system. The process of unification is accompanied by the transfer of particles (money) between the systems and explicit expression is obtained for the transferred number of particles (size of the capital).  A relation between parameters of the initial systems also found which defines the regime with zero particle transfer.


\end{abstract}

\section{Introduction}
\label{intro}
Phenomenological parallels between thermodynamic and statistical properties of systems of physical particles and economic systems are currently well known \cite{Saslow}-\cite{Maslov2011}. An analogue of identical physical particles in economics are banknotes of the same value, and the temperature of a physical system is analogous to the turnover rate in economics. The problem of interaction of two systems consisting of equal identical elements when these systems are united is very important for both physics and economics. The following discussion will consist of parallel consideration of the problem in physics and economics. The solution of the problem about the interaction of two systems in economics makes it possible to predict to what degree the corresponding collaboration is profitable.

As a model we consider two systems of equal elements (particles or banknotes) with different fractal dimensions $D_1$ and $D_2$. It is convenient to deal with
half-dimensions $\alpha=D/2$ and assume that $\alpha_1 < \alpha_2$. Each of the systems is characterized by temperature $T$, total energy $E$, entropy $S$, and the number of particles $k$. We label these parameters by subscripts 1 and 2 for the two initial systems. Similar parameters for the united system are left unlabeled. 

In economics entropy is defined as the degree of uncertainty of a state having finite or denumerable number of microstates. In other words, entropy of the system is a measure of diversity of possible microstates of the system.

Given the initial states of systems 1 and 2  we aim to describe and analyze final state of the united system. We do not analyze nonequilibrium processes in the course of unification and consider both physical and economic problems in parallel. Obviously, under unification, there is a transfer of particles (banknotes) between systems, and it is for this reason that we assume distribution of particles in the systems to be parastatistical, described by the Gentile statistics \cite{Gentile}-\cite{Dai}.

Among the possible dimensions there is an important bound $\alpha = 1$, at which a phase transition of $\lambda$-type occurs, see \cite{Kvas1}-\cite{Molotkov}. Under a phase transition of $\lambda$-type, there occurs neither a jump-like change of entropy (as under a transition of the first kind) nor a jump of the derivative of entropy with respect to temperature (as under a transition of the second kind). The graph of the temperature dependence on entropy at $\alpha = 1$ has an inflection point (see \cite{Maslov2011}, Ch. 1).

Thus, there are two significant cases: 

\begin{equation}
\label{case1}
1<\alpha_1<\alpha_2 \leq {3\over 2}
\end{equation}

\noindent where both values $\alpha_1$ and $\alpha_2$ are on the same side of the bound $\alpha = 1$, and

\begin{equation}
\label{case2}
{1\over 2} \leq\alpha_1 < 1 < \alpha_2 \leq {3\over 2}
\end{equation}
where $\alpha_1$ and $\alpha_2$ are separated by this bound. Case (\ref{case1}) corresponds to the unification of relatively close systems. 
In case (\ref{case2}) the systems differ considerably so that the resulting formulae are more complex. 

A brief outline of the first part of this work has been published in \cite{Molotkov2014}.

\section{Temperature and dimension of the united\\system}
\label{sec:1}

The parastatistical property of systems of particles under consideration makes it possible to apply Maslov'’s theorems on parastatistics 
\cite{Maslov2005}-\cite{Maslov2009}, 
on which the determination of asymptotic relations between the parameters of the systems is based. These relations contain small parastatistical parameter
$b = T^{-1}$, i.e., they are valid for large values of the temperature $T$. The form of these relations depends on the fractal dimension of the system. For

\begin{equation}
\label{aint}
1<\alpha\leq{3\over 2}
\end{equation}
the most important relations are (see \cite{Molotkov,Maslov2009,MaslovNezaikin})

\begin{equation}
\label{Easimp}
E=f(\alpha)T^{1+\alpha} ,
\end{equation}

\begin{equation}
\label{Sasimp}
S=f(\alpha)T^\alpha .
\end{equation}
Here
\begin{equation}
\label{fdef}
f(\alpha) = \alpha^2\Gamma(\alpha)\zeta(1+\alpha)
\end{equation}

$\Gamma(x)$ is the gamma-function and $\zeta(x)$ is the Riemann zeta-function. On the interval

\begin{equation}
\label{int1232}
{1\over 2} \leq\alpha\leq{3\over 2}
\end{equation}
the function $f(\alpha)$ is monotonically increasing. 
At
\begin{equation}
\label{int121}
{1\over 2} \leq\alpha < 1
\end{equation}
expression (\ref{Easimp}) for the total energy remains valid, but the expression for the entropy becomes more complicated. Calculating the entropy as the logarithm of the
number of all possible cases on the basis of methods of quantum statistic, we obtain a representation of the entropy as the sum of three terms, see \cite{Molotkov,Maslov2009,MaslovNezaikin}.
For different ranges of variation of the dimension $\alpha$, different terms of this sum turn out to prevail. 

Using formulae from \cite{Molotkov}, we can conclude that, in interval (\ref{int121}), we have
\begin{equation}
\label{Sasimp2}
S = g(\alpha) T ^{1+\alpha\over\alpha} .
\end{equation}

For the continuity of the entropy at $\alpha = 1$, it is required that $g(1) = f(1) = \zeta(2) = \pi^2/6$. Farther from the point $\alpha = 1$, 
i.e., for $1/2< \alpha < 1 - \delta$ , $\delta > 0$, it follows from relations of \cite{Dai} that

\begin{equation}
\label{gdef}
g(\alpha) = \left[  \int_0^\infty \left({1\over x} - {1\over e^x-1}\right) dx^\alpha \right]^{1/\alpha} \cong ({\alpha\over 1-\alpha})^{1/\alpha} .
\end{equation}

Relations (\ref{Sasimp}) and (\ref{Sasimp2}) on interval (\ref{int1232}) determine the entropy as a monotonically increasing function of the temperature, as it must be. 
Moreover, the graph of $S(T)$, as mentioned above, indeed has an inflection point at $\alpha = 1$.

We now proceed with the study of characteristics of the united system. 
We begin with the simpler case (\ref{case1}) and use the additivity of energy and entropy under the unification of the systems and relations (\ref{Easimp}) and (\ref{Sasimp}). 
The additivity of E and S gives

\begin{equation}
\label{f1}
f(\alpha_1)T_1^{1+\alpha_1} + f(\alpha_2)T_2^{1+\alpha_2} = 2f(\alpha)T^{1+\alpha} ,
\end{equation}

\begin{equation}
\label{f2}
f(\alpha_1)T_1^{\alpha_1} + f(\alpha_2)T_2^{\alpha_2} = 2f(\alpha)T^{\alpha} .
\end{equation}

In these relations, the resulting half-dimension $\alpha$, as well as temperature $T$ is unknown.
Multiplier 2 in (\ref{f1}) and (\ref{f2}) makes it possible to apply these equalities also for $\alpha_1=\alpha_2$ and $T_1=T_2$.

On the basis of (\ref{f1}) and (\ref{f2}), we obtain

\begin{equation}
\label{fa1fa2}
{ f(\alpha_2)\over f(\alpha_1) } = {T_1^{\alpha_1} \over T_2^{\alpha_2}} \cdot {{T_1-T} \over {T-T_2}} .
\end{equation}

Since the function $f(\alpha)$ is monotonically increasing and the left-hand side of (\ref{f1}) is positive, it follows that, among 
all possible relations between $T_1$, $T_2$, and $T$, the only possible one is

\begin{equation}
\label{T1T2}
T_1>T>T_2 .
\end{equation}

\noindent All other cases contradict (\ref{fa1fa2}). Similarly, it follows from relations (\ref{gdef})-–(\ref{fa1fa2}) that

\begin{equation}
\label{a1a2}
\alpha_1<\alpha<\alpha_2 .
\end{equation}

Thus, both the temperature and the fractal dimension of the united system are intermediate with respect to corresponding parameters of the initial systems. Yet
another important conclusion is that system with smaller dimension has higher temperature, or, in other words, an economic system with smaller dimension
is characterized by higher capital turnover rate. According to Landau theory \cite{LandauLifshits}, a system with higher temperature (and smaller dimension) possesses higher symmetry. In economics, higher symmetry presumably corresponds to a higher level of economic diversification. 

Along with inequalities (\ref{T1T2}) and (\ref{a1a2}), exact values of the characteristics $T$ and $\alpha$ of the united system can be obtained. On the basis of (\ref{f1}), after cumbersome (although elementary) calculations, we find

\begin{equation}
\label{Tmean}
T = {1\over 2} (T_1+T_2) .
\end{equation}

\noindent Similarly, using (\ref{f2}) and (\ref{Tmean}), we obtain

\begin{equation}
\label{alphamean}
\alpha = {1\over 2} (\alpha_1+\alpha_2) .
\end{equation}

In case (\ref{case1}), the temperature $T$ of the united system equals the arithmetic mean of the temperatures of systems 1 and 2, 
and the dimension $\alpha$ of the united system equals the arithmetic mean of the initial dimensions.

\section{Case of systems dimensions being separated by the phase transition point. Entropy of the united system}
\label{sec:3}

We proceed to case (\ref{case2}). Relation (\ref{f1}), which describes the additivity of energy, remains valid.
Relation expressing the additivity of entropy takes the form

\begin{equation}
\label{EntAdd}
\left({\alpha_1\over 1-\alpha_1}\right)^{1\over \alpha_1} T_1^{  {1+\alpha_1} \over \alpha_1 } + f(\alpha_2) T_2^{\alpha_2} = h(\alpha) T^\alpha .
\end{equation}

\noindent Here function $h(\alpha)$ and quantities $\alpha$, $T$, $T1$, and $T2$ are unknown, and it is only required to estimate the parameters $T$ and $\alpha$ of the united system.
First of all, we need to obtain for the case (\ref{case1}) the generalized relation (\ref{Tmean}) between $T$ and arithmetic mean $ (T_1+T_2)/2 = \tau $. Most significant in (\ref{EntAdd}) are temperature power terms. 
It should also be noted that $\alpha_2>{1+1/\alpha_1}$  cannot be true as it would then follow that $\alpha_2>2$, but this contradicts (2). 
For 

\begin{equation}
\alpha_2 < 1 + {1\over \alpha_1} 
\end{equation}

\noindent taking into account (\ref{EntAdd}), in the highest order of magnitude we obtain

\begin{equation}
\left(  { \alpha\over 1-\alpha_1}\right)^{1\over{1-\alpha_1}}\tau^{1+{1\over \alpha_1} } = h(\alpha) T^\alpha .
\end{equation}

\noindent From this it follows that

\begin{equation}
\label{Ttau}
T=\tau^{1+\alpha_1\over \alpha\alpha_1}
\end{equation}

\noindent and

\begin{equation}
\label{halpha}
h(\alpha)=\left( {\alpha\over{1+\alpha_1}} \right)^{1\over{1-\alpha_1}} .
\end{equation}

\noindent As $(1+\alpha_1)/\alpha\alpha_1 < 1$, then $T<\tau$. From (\ref{Ttau}) follows an estimation 

\begin{equation}
\label{ttau}
T-\tau   \approx - {1+\alpha_1\over \alpha\alpha_1} \ln \tau < 0
\end{equation}

\noindent which defines deviation of the final temperature (turnover rate) from the arithmetic mean of the initial temperatures. 
Inequalities (\ref{T1T2}) thus hold both in case (1) and in case(2). Final temperature of the united system is thus lower than the 
arithmetic mean  temperature of systems 1 and 2. 

Transformations similar to those performed above also prove the validity of inequalities (\ref{a1a2}). 
In a similar way, we arrive at the conclusion that the dimension of the united system is higher than the arithmetic mean 
of the dimensions of the initial systems.

The obtained formulae define entropy of the united and initial systems. Entropies of the initial systems 1 and 2 are 
represented by the left-hand side of the equation (\ref{EntAdd}). 
Entropy of the united system equals to the right-hand side of equation (\ref{EntAdd}) where $h(\alpha)$  is given by (\ref{halpha}). 

Equation (\ref{EntAdd}) shows that contributions of systems 1 and 2 into the entropy of united system are quite different. 
When $\alpha_1<1$  major contribution is given by the more complex (according to L.D. Landau theory) system 1. 
It allows for the much greater number of  microstates which means higher entropy for this system.

\section{Particle and capital transfer as the result of unification}
\label{sec:4}

The apparatus presented above also makes it possible to quantitatively estimate the displacement of
particles and money under unification. Suppose that condition (\ref{case1}) holds.
By analogy with relations (\ref{Easimp}) and (\ref{Sasimp}), we can derive the relation (see~\cite{Molotkov})

\begin{equation}
\label{kkappa}
k\kappa = T[(\alpha - 1)\ln k + \alpha\ln \kappa]
\end{equation}

\noindent between the number of particles, temperature and chemical potential $\mu = -\kappa, \kappa>0$ .

In economics, chemical potential ("nominal interest rate") is the ratio of the small increase in the amount of money in a system after infinitesimal intervention,  to the value of this intervention.

The left-hand side of (\ref{kkappa}) represents the Gibbs potential $G=k\mu=-k\kappa$, taken with the opposite sign. Thus, equation (\ref{kkappa}) can be rewritten as follows:

\begin{equation}
\label{Gibbs}
|G| = T(\alpha\ln|G|-\ln k) .
\end{equation}

Main difficulty in using (\ref{kkappa}) comes from the fact, that universal dependence of $\kappa$ on $\alpha$ is unknown. Special case $\alpha=3/2$ was   considered in \cite{Molotkov2014}.  For this case in \cite{Kvas2}, Ch. 2, it was found, that

\begin{equation}
\label{kappat}
\kappa = {3\over 2} T \ln T + O(T)
\end{equation}

Relation (\ref{kappat}) made it possible in \cite{Molotkov2014} to eliminate the chemical potential and turn (\ref{kkappa}) into a relation between 
the number of particles and temperature for $\alpha=3/2$. In this special case it turned out that the number of particles in the system with higher dimension
decreases after unification, which means an outflow of capital from this system into the system with lower dimension.

In order to estimate particle and capital transfer in a more general case we now consider unification of two systems 1 and 2, satisfying (\ref{case1}) and having 
close dimensions and temperatures. Let $T_1, \alpha_1, k_1, \kappa_1$ be parameters of system 1 before the unification. After the unification, in the system 1 domain, according to (\ref{T1T2}) -- (\ref{alphamean}) we obtain

$$
\begin{tabular}{ll}
$T=T_1-\Theta ,$  & $\alpha=\alpha_1+\Delta=\alpha_2-\Delta$ \\
$k=k_1+\lambda ,$  & $\kappa =\kappa_1+\xi$ \\
\end{tabular}
$$

\noindent Here additional terms $\Theta,\Delta,\lambda,\xi$ having small absolute value, $\Theta > 0, \Delta>0$, signs of $\lambda$ and $\xi$ are unknown.

How small changes of temperature depend on small changes in fractal dimensions has been established earlier in \cite{Molotkov} :

\begin{equation}
\label{Tdim}
(T_1-T)\alpha_1 = (\alpha-\alpha_1)T_1 q_1 ,
\end{equation}

\begin{equation}
\label{q1}
q_1 \equiv {f'(\alpha_1)\over f(\alpha_1)} + \ln T_1 .
\end{equation}

\noindent Function $f(\alpha)$ is defined by (\ref{fdef}) and is monotonically increasing in (\ref{case1}). 

\noindent From (\ref{kkappa}) we obtain

\begin{equation}
\label{kkappa1}
k_1\kappa_1 = T_1\ln{(k_1\kappa_1)^{\alpha_1}\over k_1} ,
\end{equation}

\begin{equation}
\label{kkappa1plus}
k_1\kappa_1+\lambda\kappa_1+\xi k_1 = (T_1-\Theta)\ln {(k_1\kappa_1+\lambda \kappa_1+\xi k_1)^{\alpha_1+\Delta}\over {k_1+\lambda}} .
\end{equation}

\noindent By subtracting (\ref{q1}) from (\ref{kkappa1plus}) all terms that do not include small coefficients mutually cancell and after taking into account (\ref{Tdim}), result takes the form

$$
\lambda H_1 + \xi H_2 = \Delta H_3 .
$$

\noindent where

$$
H_1 \equiv {1\over k_1} \left[ {k_1\kappa_1\over T_1} - (\alpha_1 - 1) \right]   ,      H_2 \equiv {1\over \kappa_1} \left[ {k_1\kappa_1\over T_1} - \alpha_1) , \right]  
$$
$$
H_3 \equiv \left( 1 - {\alpha_1 - 1 \over \alpha_1} q_1  \right) \ln k_1 + (1-q_1)\ln \kappa_1
$$

\noindent Or

\begin{equation}
\label{lambda}
\lambda = {H_3\Delta - H_2\xi \over H_1}
\end{equation}

The sign of $\lambda = k - k_1$ points to the direction of money or capital transfer. Two terms in the (\ref{lambda}) numerator represent two concurrent processes going on during the unification of two systems. Increase of temperature along with the decrease of dimension, see (\ref{Tdim}), described by the term $H_3\Delta$,
leads to increase in the number of particles (capital). Decrease of chemical potential, in contrast, decreases the number of particles. When numerator (\ref{lambda}) equals zero, number of particles does not change, the transfer is absent.

\section{Resulting formulae for the particle transfer. Size of the transferred capital}
\label{sec:5}

We can now try to determine the condition, under which the transfer of particles vanishes.  According to (\ref{lambda}) this process stops when

\begin{equation}
\label{stopcond}
\left( {k_1\over T_1}-{\alpha_1\over\kappa_1}  \right) \xi = -q_1\left( {\alpha_1-1\over\alpha_1} \ln k_1 + \ln \kappa_1  \right)\Delta .
\end{equation}

\noindent In (\ref{stopcond}) it was taken into account that $q_1 = f'(\alpha_1)/f(\alpha)+\ln T_1$ is much greater than unity which follows from 
calculations of $f(\alpha)$.

From (\ref{fa1fa2}) and (\ref{Tdim}) follows the relation

\begin{equation}
\label{q1da}
q_1d\alpha = {\alpha_1\over T_1}dT
\end{equation}

\noindent between changes in dimension and temperature. Uisng the Gibbs potential G we can rewrite (\ref{stopcond}) as follows

\begin{equation}
\label{G1at}
{|G_1|-\alpha_1T_1 \over \kappa }d\kappa = -\left( \alpha_1\ln|G_1| - \ln k_1 \right)dT .
\end{equation}

In order to further simplify (\ref{G1at}) we can use a well-known thermodynamical relation \cite{Kvas1,LandauLifshits} for the entropy, enthalpy and Gibbs potential. The united system (at  the final stage of unification process) can be considered closed and at equilibrium so that its temperature and pressure are constant and enthalpy is at its minimum. In this case enthalpy can be neglected and we thus obtain the Gibbs-Duhem equation.

\begin{equation}
\label{GibbsDuhem}
dG = -TdS
\end{equation}
 
\noindent Using (\ref{Sasimp}), (\ref{q1da}) and considering $S$ being function of both $T$ and $\alpha$, we find that

$$
dS = 2\alpha S T^{-1}dT
$$
and

\begin{equation}
\label{dG}
dG = -kdk=-2\alpha SdT
\end{equation}

\noindent When temperature decreases (and, hence, dimension increases) $\kappa$ decreases as well. 

Comparing (\ref{G1at}) and (\ref{dG}) and taking into account that $ST = k\kappa$, we obtain:

$$
2\alpha_1S_1-2\alpha_1^2 = \ln k_1 - \alpha_1\ln(T_1S_1)
$$
 
\noindent As $S_1=f(\alpha_1)T_1^{\alpha_1}$ is much greater than $\alpha$ and also considerably greater than $\ln T_1$ and $\ln S_1$ to the 
highest order the relation of vanishing transfer takes simple form

\begin{equation}
\label{noflow}
\alpha_1f(\alpha_1)T_1^{\alpha_1} = \ln k_1 .
\end{equation}

\noindent The condition (\ref{noflow}) is a special relation between parameters $\alpha_1, T_1$ and $k_1$.  Also, in (\ref{lambda}) $\lambda=k-k_1=0$. It should also be noted, that if $\alpha$ satisfies (\ref{case1}), then, as numerical analysis show

$$
\label{afa}
1,5 \leq \alpha f(\alpha) \leq 3,4
$$

\noindent If, instead of (\ref{noflow}),

\begin{equation}
\label{noflow1}
\alpha_1f(\alpha_1)T_1^{\alpha_1} > \ln k_1
\end{equation}

\noindent (for example, due to increase in temperature $T_1$), then $\lambda>0, k>k_1$ and number of particles (size of capital) in system 1 as the result of unification increases at the expense of system 2 with greater dimension. It is clear that increase of $T_1$ (rate of capital turnover) increases chances for system 1 to see additional growth of its capital after unification.

From (\ref{noflow1}) it is clear that if sign is changed to the opposite, then transfer of particles goes in the opposite direction -- number of particles in system 1 decreases while system 2 acquires these particles. 

We can now proceed with the calculation of the size of capital transfer due to the unification. Let us assume that (\ref{noflow1}) holds.  Using (\ref{lambda}) we obtain

$$
dk = {-q_1T_1\left({{\alpha_1-1}\over\alpha_1}  \ln k_1+\ln\kappa_1 \right)d\alpha  +{1\over\kappa_1} (|G_1|-\alpha_1 T_1) d\kappa \over {1\over k_1}\left(|G_1|-(\alpha_1-1)T_1\right)}
$$

\noindent which links $dk, d\alpha$ and $d\kappa$. Uisng (\ref{q1da}), (\ref{GibbsDuhem}) and (\ref{dG}) we arrive at

\begin{equation}
\label{dkdT}
dk = {-\alpha \ln|G_1|-\ln k_1+2\alpha_1S_1-2\alpha_1^2\over {1\over k_1}\left(T_1S_1-(\alpha_1-1)T_1 \right)}dT .
\end{equation}

\noindent  We now again take into account that $S_1$ is much greater than $\alpha_1, \ln T_1,\ln S_1$. After some straightforward transformations and elementary integration  we obtain, to the highest order

\begin{equation}
\label{k-k1}
k-k_1=k_1(T_1-T) {2\alpha_1f(\alpha_1)T_1^{\alpha_1}+(\alpha_1-1)\ln k_1\over f(\alpha_1) T_1^{\alpha_1+1}} .
\end{equation}

\noindent Here $k-k_1>0$ - increase in the number of particles (size of capital) in system 1, $T$ is the temperature of the united system.  From (\ref{Tmean}) it follows, that $T_1 -T > 0$.

Using dimensionless ratio of the capital surplus $k-k_1$ to its initial value $k_1$, (\ref{k-k1}) acquires a more clear form. After transformations, 
using (\ref{Sasimp}), (\ref{kkappa}) and (\ref{GibbsDuhem})

\begin{equation}
\label{k-k1k1}
{{k-k_1}\over k_1}= {  {T_1-T}\over T_1 }  { {(2\alpha_1+1)(\alpha_1-1)\ln k_1 + 2\alpha_1^2\ln \kappa_1}\over{(\alpha_1-1)}\ln k_1+\alpha_1\ln \kappa_1} > 
2\alpha _1 { {T_1-T}\over T_1} .
\end{equation}

\noindent From this it follows that relative increment of capital (\ref{k-k1k1}) in system 1 is defined by the relative decrease of capital turnover rate in this system after unification, and its dimension.

\section{Conclusions}
\label{sec:6}

In physics, difference between dimensions $\alpha_1$ and $\alpha_2$ of systems of identical particles implies difference in pressures or densities of these systems. In the system with lower dimension $\alpha$, temperature $T$ is higher. Temperature of the united system either equals the arithmetic mean of  the temperatures of initial systems (case (\ref{case1}), see (\ref{Tmean})) or is lower than this mean value (in case (\ref{case2}), see (\ref{ttau})). 

In economics, $T$ has the meaning of the rate of capital turnover (see \cite{Saslow}, \cite{Maslov2012}).  System with lower dimension is characterized by the higher rate of capital turnover. This, presumably, means that the system possesses higher level of economic diversification, lower "economic viscosity".  In case (\ref{case2}) dimension of the united system is higher than mean dimension of the initial systems. This means that the united system possesses a somewhat lower level of economic diversification and lower turnover rate than corresponding mean value of the initial systems. 

Our analysis of the particle and capital transfer for the systems under unification is based on (\ref{kkappa}). We have considered a simple case (\ref{case1}). Our consideration of two systems having comparable dimensions has allowed us to arrive at basic relation (\ref{lambda}) for the number of particles and capital transfer in this case. A condition (\ref{noflow}) was found when particle (capital) transfer vanishes. For the general case of two systems, capital transfer is found to be described by  (\ref{k-k1}), (\ref{k-k1k1}).  Also, (\ref{k-k1k1}) includes a simple and clear estimation of the size of this capital. Capital transfer calculations take into account simultaneous  changes in temperature, dimension and chemical potential of the systems under consideration.





\end{document}